\documentclass[conference]{IEEEtran}
\IEEEoverridecommandlockouts
\usepackage{cite}
\usepackage{amsmath,amssymb,amsfonts}
\usepackage{algorithmic}

\usepackage{caption}
\usepackage{subcaption}
\usepackage{xcolor}
\usepackage{bm}

\usepackage{hyperref}
\usepackage[ruled]{algorithm2e}
\usepackage{multirow}
\usepackage{caption}
\usepackage{subcaption}
\usepackage{graphicx}

\def\BibTeX{{\rm B\kern-.05em{\sc i\kern-.025em b}\kern-.08em
    T\kern-.1667em\lower.7ex\hbox{E}\kern-.125emX}}
\begin{document}

\title{Efficient Sampling for Data-Driven Frequency Stability Constraint via Forward-Mode Automatic Differentiation
}

\author{\IEEEauthorblockN{Wangkun Xu, Qian Chen, Pudong Ge, Zhongda Chu, Fei Teng\IEEEauthorblockA{\textit{Electrical and Electronic Engineering, Imperial College London, London, UK} \\
\{wangkun.xu18, qian.chen19, pudong.ge19, z.chu18, f.teng\}@imperial.ac.uk}
}}

\maketitle

\begin{abstract}
Encoding frequency stability constraints in the operation problem is challenging due to its complex dynamics. Recently, data-driven approaches have been proposed to learn the stability criteria offline with the trained model embedded as a constraint of online optimization. However, random sampling of stationary operation points is less efficient in generating balanced stable and unstable samples. Meanwhile, the performance of such a model is strongly dependent on the quality of the training dataset. Observing this research gap, we propose a gradient-based data generation method via forward-mode automatic differentiation. In this method, the original dynamic system is augmented with new states that represent the dynamic of sensitivities of the original states, which can be solved by invoking \textit{any} ODE solver for a \textit{single} time. To compensate for the contradiction between the gradient of various frequency stability criteria, gradient surgery is proposed by projecting the gradient on the normal plane of the other. In the end, we demonstrate the superior performance of the proposed sampling algorithm, compared with the unrolling differentiation and finite difference. All codes are available at \url{https://github.com/xuwkk/frequency_sample_ad}.
\end{abstract}


\section{Introduction}

Accurate and efficient stability assessment is essential for the safe operation of the power system \cite{gao2022lyapunov}. With the high penetration of renewables, the inertia of the system decreases, making it more sensitive to disturbance and vulnerable to cyber intrusions \cite{chu2023mitigating}. As a result, many researchers have tried to account for security assessments at the operation stage. Depending on the dynamic model of the power systems and the type of stabilities, different formulations of security constraints are encoded into tractable optimizations. For small-signal stability, the dominant eigenvalue of the system state matrix can be directly used as the security index \cite{li2017stability}. The eigenvalue sensitivities can also serve as gradients in sequential quadratic programming and the interior point method. However, frequency stability assessment is rather complicated, which is dictated by post-fault frequency evolution obtained from solving an ordinary differential equation (ODE) describing the dynamics of the power systems. An extensive research effort has been conducted to derive closed-form frequency constraints respecting the prescribed stability criteria, such as the maximum admissible values for frequency steady-state (SS) deviation, frequency nadir, and maximum rate of change of frequency (RoCoF) \cite{badesa2019simultaneous, zhang2020modeling}. In these papers, the authors have compromised the dynamics of frequency for the sake of their analytical formulations. In contrast, the time-domain frequency response would preserve all dynamic information. However, encoding and solving ODEs present a large obstacle in the optimization problem, which is the main incentive behind the wide use of data-driven approaches in security assessments in power systems \cite{bellizio2023transition, zhang2023data, cremer2019optimization}. 

The performance of the data-driven model is strongly dependent on the quality of the training dataset \cite{cremer2019optimized}. First, the large number of balanced stable and unstable operating points is essential for supervised training. Second, it is desirable to construct the informative dataset so that the operating points are around the true decision boundary \cite{li2017stability}. The lack of gradient (or local sensitivity) information hinders the efficient sampling of new data from the existing dataset. Due to this bottleneck, most existing literature still applies the random sampling method to the frequency security problem \cite{li2023optimal}, which is less efficient and uncontrollable. 

This paper aims to fulfill this research gap using the gradient-based sampling method. To efficiently obtain the Jacobian of the various frequency criteria with respect to the stationary operating points, an augmented system of ODEs is constructed, including the original states of the system and its associated tangents. Notably, the tangent states form linear system that can be solved by any ODE solver, together with the original ones. To overcome the contradiction between the gradients of different stability criteria, gradient surgery is implemented to eliminate the contradictory component. 

\section{Data-Driven Frequency Stability Assessment}

\subsection{Dynamic Model}

Although our algorithm is general and has the potential to sample data from dynamic power system models with various stability criteria, we pay special attention to frequency stability under power system load frequency control (LFC) with adaptive synthetic inertia provision. For example, the wind turbine is operated in grid-forming (GFM) mode, which can be modeled by nonlinear ODEs. A detailed deviation of this section can be found in \cite{markovic2018fast}.

To start, a second-order system composed of generator swing equation and turbine governor is given as \cite{saadat1999power},
\begin{subequations}\label{eq:swing_governor}
    \begin{equation}\label{eq:swing_equation}
        M \cdot \dot{\omega} = \Delta P + q -D \cdot \omega
    \end{equation}
    \begin{equation}
        \tau \cdot \dot{q} =-r^{-1} \cdot \omega - q
    \end{equation}
\end{subequations}
where $M$ and $D$ are the normalized inertia and damping constants, respectively; $\omega$ is the change of the frequency; $q$ and $\Delta P$ are the perturbed mechanical and electrical power (the load disturbance), respectively; $r$ and $\tau$ are the droop gain and time constant of the governor, respectively.

From \eqref{eq:swing_governor} and by assuming that the constant step change disturbance, the second-order differential equation of $\omega$ is given as
\begin{equation}\label{eq:second_order}
        \ddot{\omega}=-\left(\frac{D}{M}+\frac{1}{\tau}\right) \dot{\omega} - \left(\frac{1}{r \tau M} + \frac{D}{\tau M}\right) \omega
         +\frac{\Delta P}{\tau M}
\end{equation}

Let $x = [\omega, \dot{\omega}]^T$ be the state. A nonlinear ODE can be constructed by the following state space model
\begin{subequations}\label{eq:open_loop}
    \begin{equation}
        \dot{x}_1 = x_2
    \end{equation}
    \begin{equation}
        \dot{x}_2 = -\left(\frac{1}{r \tau M}+\frac{D}{\tau M}\right)x_1 -\left(\frac{D}{M}+\frac{1}{\tau}\right) x_2 + \frac{\Delta P}{\tau M}
    \end{equation}
\end{subequations}

Converter in the virtual synchronous mode (VSM) is considered such that inertia $M$ and damping $D$ are the control inputs to the system by the following full state-feedback control \cite{markovic2019lqr},
\begin{equation}\label{eq:state_feedback}
    \begin{bmatrix}
        M \\ D     
    \end{bmatrix} = \begin{bmatrix}
        M_0 \\ D_0
    \end{bmatrix} - \begin{bmatrix}
         K_{11} &  K_{12} \\
         K_{21} & K_{22}
    \end{bmatrix} \begin{bmatrix}
        x_1 \\ x_2
    \end{bmatrix}
\end{equation}
where $M_0$ is the nominal inertia of the synchronous generator (SG); $D_0$ is the nominal damping coefficient; $K$s are the proportional control gains. Combining \eqref{eq:open_loop} and \eqref{eq:state_feedback} results in the closed-loop dynamic of the LFC model with adaptive virtual inertia provision:
\begin{equation}\label{eq:closed_loop}
    \eqref{eq:open_loop}, \; \eqref{eq:state_feedback}, \; x(t=0) = [0, \dot{\omega}(t = 0^+)]^T
\end{equation}
where
\begin{equation}\label{eq:init_omega_dot}
    \dot{\omega}(t=0^+) = \frac{M_0 - \sqrt{M_0^2 - 4K_{12}\Delta P}}{2K_{12}}
\end{equation}
A derivation can be found in the Appendix \ref{sec:initial_condition}. The permissible range of $K_{12}$ can be derived from \eqref{eq:init_omega_dot} and it is also easy to show that $\dot{\omega}(t=0)\leq 0$ if $\Delta P < 0$ for any $K_{12}$.

\subsection{Frequency Stability}

There are three main frequency stability criteria, namely, the maximum instantaneous RoCoF $\tau_{rocof}$, maximum steady-state frequency deviation $\tau_{ss}$, and the frequency nadir $\tau_{nadir}$. In the following discussion, the simulation time is considered as $t\in[0,T]$ and the critical time steps to achieve the maximum RoCoF, nadir and SS are $t_{rocof}$, $t_{nadir}$, and $t_{ss} = T$, respectively. Note that the dynamic \eqref{eq:closed_loop} is non-linear, so it lacks an analytic time domain solution such as in \cite{markovic2018fast, markovic2019lqr} and the determination of the critical time steps requires solving \eqref{eq:closed_loop}. Consequently, directly encoding the stability constraints into an optimization requires explicit formulation on the stability criteria. For the linearized version of \eqref{eq:closed_loop}, it is possible to derive the analytic forms under certain assumptions, and their dependence on $M$, $r$, and $D$ can be exactly found \cite{cui2024control}, although it is still difficult to directly encode them into an optimization problem.

Due to the complexity of directly encoding non-linear ODEs or the linearized time-domain solution into an optimization problem, data-driven approaches have been developed to learn a more tractable relationship (such as a convex or mixed-integer linear function) between the operating condition (e.g. the parameters of the ODE; in our case, the $K$) to three criteria. 

\section{Gradient-based Sampling under Forward-Mode Automatic Differentiation}

In this paper, we study how to efficiently construct the dataset $\theta = [K_{11}, K_{12}, K_{21}, K_{22}]^T \times y \in \mathcal{D} = \mathcal{X}\times\mathcal{Y}$ under predefined disturbances $\Delta P$. $y\in\{0,1\}$ with 0 represents a stable operation such that all three criteria are satisfied. We assume that there exists a dataset $\mathcal{D}_0$ of historical randomly-sampled operating conditions. The dataset is either insufficient in size or strongly biased to one class, and our goal is to efficiently increase the size of the dataset with \textbf{controllable} labeling. 

In general, denote the non-linear closed-loop dynamic \eqref{eq:closed_loop} as
\begin{equation}\label{eq:ode}
    \frac{d x(t)}{d t} = f(x(t), \theta), \quad x(t=0) = x_0
\end{equation}
When the stability label $y'$ for a sample $\theta'$ equals 0, an efficient sampling strategy may be to find an unstable sample $\bar{\theta}'$ which is the `closest' to the original sample $\theta'$. Intuitively, such a sampling method may construct the dataset that can better represent the true stability boundary and give benefits in training the data-driven assessor with less data. One direct consideration is to use the local sensitivity of $\theta$ to iteratively approach the stability boundary. Similar consideration has been adopted in sampling data for assessing small-signal stability \cite{li2017stability}. However, as neither $x(t)$ has a closed form nor the critical $t$s can be specified in advance, the sensitivity-based method needs to be reconsidered for frequency stability.  

\subsection{Forward-Mode Automatic Differentiation}

Our goal is to find the sensitivities of $x$ to $\theta$, e.g., the Jacobian matrix $\frac{\partial x(t)}{\partial \theta}, t\in \{t_{ss}, t_{rocof}, t_{nadir}\}$, which can fully represent the search direction of $\theta$. 

We propose to efficiently find the sensitivity $\frac{\partial x(t)}{\partial \theta}$ by automatic differentiation in the forward mode (FMAD). The basic idea is to associate the initial value of the ODE with unit tangent vectors and solve an augmented ODE for both the original ODE of $x$ and the new ODE of $\frac{\partial x(t)}{\partial \theta}$. 

Since $\theta\in\mathbb{R}^4$, define four new states $\hat{x}_i = \frac{\partial x}{\partial \theta_i} = \frac{\partial x}{\partial \theta}e_i,i=1,\cdots,4$ where $e_i\in\mathbb{R}^4$ is a unit vector with the $i$-th entry equal to one. The augmented ODE with FMAD can be written as

\begin{subequations}\label{eq:fmad}
    \begin{equation}
        \frac{d x(t)}{d t} = f(x(t), \theta)
    \end{equation}
    \begin{equation}\label{eq:augment_ode}
        \frac{d \hat{x}_i(t)}{d t} = \frac{\partial f}{\partial x} \hat{x}_i(t) + \frac{\partial f}{\partial \theta}e_i, \quad i=1,\cdots,4
    \end{equation}
\end{subequations}
with initial conditions
\begin{subequations}\label{eq:fmad_initial}
    \begin{equation}
        x(t=0^+) = x_0
    \end{equation}
    \begin{equation}
        \hat{x}_2(t=0^+) = \left[0, \frac{\partial \dot{\omega}(t=0^+)}{\partial K_{12}}\right]^T
    \end{equation}
    \begin{equation}
        \hat{x}_i(t=0^+) = 0, \quad i=1,3,4
    \end{equation}
\end{subequations}

A derivation can be found in Appendix \ref{sec:derive_fmad}. Note that the initial value of $\hat{x}_2$ is also a function of $K$ and a \textbf{single} call of \textbf{any} ODE solver to \eqref{eq:fmad} results in the solution $x(t)$ to the original ODE \eqref{eq:ode} and all the required sensitivities $\frac{\partial x(t)}{\partial \theta_i},i=1,\cdots,4, t\in[0,T]$ are available. Compared to the original ODE \eqref{eq:ode}, the FMAD augmented ODE \eqref{eq:augment_ode} contains four linear inhomogeneous differential equations with zero initial values.

\subsection{Gradient Surgery}

As discussed above, the Jacobians $\frac{\partial x(t)}{\partial \theta}, t\in\{t_{ss},t_{rocof},t_{nadir}\}$ already contain all the information about the next search direction. Using RoCoF as an example, $t_{rocof}$ can be found by directly evaluating $\arg\max_t |x_2(t)|$. If $|x_2(t_{rocof})| > \tau_{rocof}$ and the target is to stabilize this sample, then $\theta$ can be updated as $\theta^+ = \theta + \alpha\cdot g_{rocof}(\theta) $ where $g_{rocof}(\theta) = -\frac{\partial |x_2(t_{rocof})|}{\partial \theta} $  and $\alpha$ is a predefined step size. Similarly, steady-state frequency and frequency nadir can be found as $g_{ss}(\theta) = \pm \frac{\partial |x_1(t_{ss})|}{\partial \theta}$ and $g_{nadir}(\theta) = \pm \frac{\partial |x_1(t_{nadir})|}{\partial \theta}$, respectively. The sign $\pm$ depends on the direction of search (e.g. negative sign when searching from `unstable' to `stable' samples and positive sign when searching from `stable' to `unstable' sample). In the following discussion, it is assumed that the proper signs have been included in the sensitivities $g$s.

During the update, one might encounter the situation where the norms of three sensitivities are significantly different and the directions are opposite to each other. Therefore, a naive linear combination may result in the direction that is only effective in updating one criterion. To avoid the contradiction of different stability criteria, we adopt the method called \textit{gradient surgery} in multi-task learning \cite{yu2020gradient}. For each gradient $g$ in $\mathcal{G} = \{g_{rocof},g_{ss},g_{nadir}\}$, we iteratively project $g$ onto the normal plane of each of the remaining gradients, if the cosine similarity between the two gradients is negative. The pseudo-code of gradient surgery can be found in Algorithm \ref{alg:gradient_surgery}.

\begin{algorithm}
    \SetKwInOut{Input}{Input}
    \SetKwInOut{Output}{Output}
    \footnotesize

    \Input{set of gradients $\mathcal{G} = \{g_{rocof},g_{ss},g_{nadir}\}$}
    \Output{aggregated gradient $\hat{g}$}
    $\bar{g} = 0$
    
    \For{$g \in \mathcal{G}$}
    {
        $g^p = g$
        
        \For{$g' \in \mathcal{G}\setminus g$}
        {      
            \tcc{Project to normal space}
            \If{$(g^p)^T\cdot g' < 0$}
            {
                $g^p := g^p - \frac{(g^p)^T\cdot g^p}{\|g'\|^2} g'$
            }
        }
        $\bar{g} = \bar{g} + g^p$
    }
    \caption{Gradient Surgery}
    \label{alg:gradient_surgery}
\end{algorithm}

\subsection{Computation and Discussion}

In \eqref{eq:augment_ode}, the Jacobians $\frac{\partial f}{\partial x}$ and $\frac{\partial f}{\partial \theta}$ can be found using an analytical method or by automatic differentiation packages such as \texttt{autograd} and \texttt{PyTorch}. As Jacobines need to be iteratively updated when solving ODEs, a more efficient and robust approach is to directly find the Jacobian-Vector Products (JVPs) \cite{baydin2018automatic} of the two products on the right-hand side of \eqref{eq:fmad} similar to the FMAD mentioned above. Therefore, \eqref{eq:fmad} needs to evaluate $f(\cdot)$ at $(x,\theta)$ four times to return the ten states without calculating and storing the Jacobian matrices explicitly. We can also fully use the parallel computation power of AD libraries to simultaneously sample a batch of $\theta$ and solve \eqref{eq:augment_ode}. 

The pseudocode for the proposed sampling method is summarized in Algorithm \ref{alg:sampling}.

\begin{algorithm}
    \SetKwInOut{Input}{Input}
    \SetKwInOut{Output}{Output}
    \footnotesize
    \Input{initial dataset $\Theta_0$, system parameters $r$, $\tau$, $M_0$, $D_0$, $\Delta P$, ODESolver, batch size $b$, maximum iteration $m$, sampling rule $R$, step size $\alpha$}
    \Output{augmented dataset $\Theta$}
    $\Theta = \{\cdot\}$, $k=1$

    \For{$\{\theta^i\}^{i=1,\cdots b} \in \Theta_0$} 
    {   
        \tcc{Sample mini-batch}
        \While{$k\leq m$ and $R$ is not satisfied}
        {
            $\{x^i(t)\}_{t\in[0,T]}^{i=1,\cdots b}, \{\frac{\partial x^i(t)}{\partial \theta^i}\}_{t\in[0,T]}^{i=1,\cdots b}\xleftarrow[\text{in parallel}]{\text{ODESolver}}$ \eqref{eq:fmad} with initial condition \eqref{eq:fmad_initial} and parameters $\{\theta^i\}^{i=1,\cdots b}$ 
    
            Find the critical time steps: $\{t_{ss}\}^{i=1,\cdots b}$, $\{t_{rocof}\}^{i=1,\cdots b}$, $\{t_{nadir}\}^{i=1,\cdots b}$
    
            \tcc{Find the stability gradients}
            $\{g_{ss}^i\}^{i=1,\cdots b}$, $\{g_{rocof}^i\}^{i=1,\cdots b}$, $\{g_{nador}^i\}^{i=1,\cdots b} \leftarrow$ $\{\frac{\partial x_1^i(t_{ss})}{\partial \theta^i}\}^{i=1,\cdots b}$, $\{\frac{\partial x_2^i(t_{rocof})}{\partial \theta^i}\}^{i=1,\cdots b}$, $\{\frac{\partial x_1^i(t_{nadir})}{\partial \theta^i}\}^{i=1,\cdots b}$
    
            \tcc{Gradient surgery}
            $\{\bar{g}^i\}^{i=1,\cdots,b}\xleftarrow[]{\text{Algorithm } \ref{alg:gradient_surgery}}$ $\{g_{ss}^i\}^{i=1,\cdots b}$, $\{g_{rocof}^i\}^{i=1,\cdots b}$, $\{g_{nador}^i\}^{i=1,\cdots b}$
    
            Update $\{\theta^i\}^{i=1,\cdots b} = \{\theta^i\}^{i=1,\cdots b} + \alpha \cdot \{\bar{g}^i\}^{i=1,\cdots,b}$

            $k=k+1$
        }
        Add $\theta^i, i=1,\cdots b$ to $\Theta$
    }
    \caption{Stationary Point Sampling Using FMAD}
    \label{alg:sampling}
\end{algorithm}

\section{Case Studies}

In the case study, we consider the power system dynamic given by \eqref{eq:closed_loop} with parameters summarized in Table \ref{tab:params}. All the ODEs are solved with the Euler's method. All the experiment is running on Ubuntu 18.04 with Intel Xeon Gold 6230R CPU @ 2.10GHz.

\begin{table}[h]
    \centering
    \footnotesize
    \caption{Power System Parameters}
    \begin{tabular}{c|c||c|c}\hline
        Parameter & Value & Parameter & Value \\\hline\hline
       $r$  & 0.06 p.u. & $\tau_{ss}$ & 0.5 Hz \\\hline
       $\tau$  & 10.0 s  & $\tau_{nadir}$ & 0.8 Hz \\\hline
       $M_0$ & 6.0 s   & $\tau_{rocof}$ & 1.0 Hz \\\hline
       $D_0$ & 5.0 p.u. & $\Delta P$ & -0.12 p.u.\\\hline
    \end{tabular}
    \label{tab:params}
\end{table}

The initial dataset $\Theta_0$ of size 100 is randomly sampled from normal distribution with mean equal to 0 and std equal to 50. The augmented dataset $\Theta$ can be sampled in parallel for the entire 100 samples in $\Theta$. We also found that the steady-state deviation does not depend on the parameter $K$ for the dynamic model \eqref{eq:closed_loop}. It is possible to prove this statement using the method in \cite{saadat1999power}. However, it is not included due to the page limit. As a result, the steady-state criterion is omitted in the following study. 

\subsubsection{Two Examples}

\begin{figure}[t]
     \centering
     \begin{subfigure}[b]{0.43\textwidth}
         \centering
         \includegraphics[width=\textwidth]{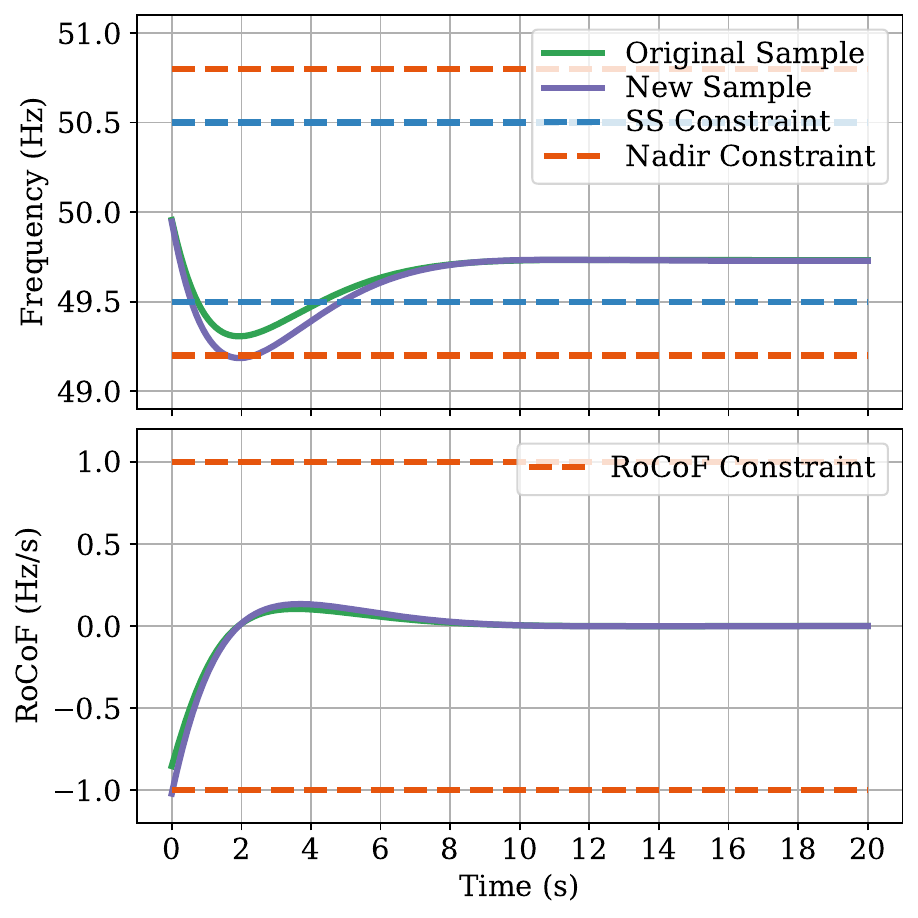}
         \caption{}
     \end{subfigure}
     \hfill
     \begin{subfigure}[b]{0.43\textwidth}
         \centering
         \includegraphics[width=\textwidth]{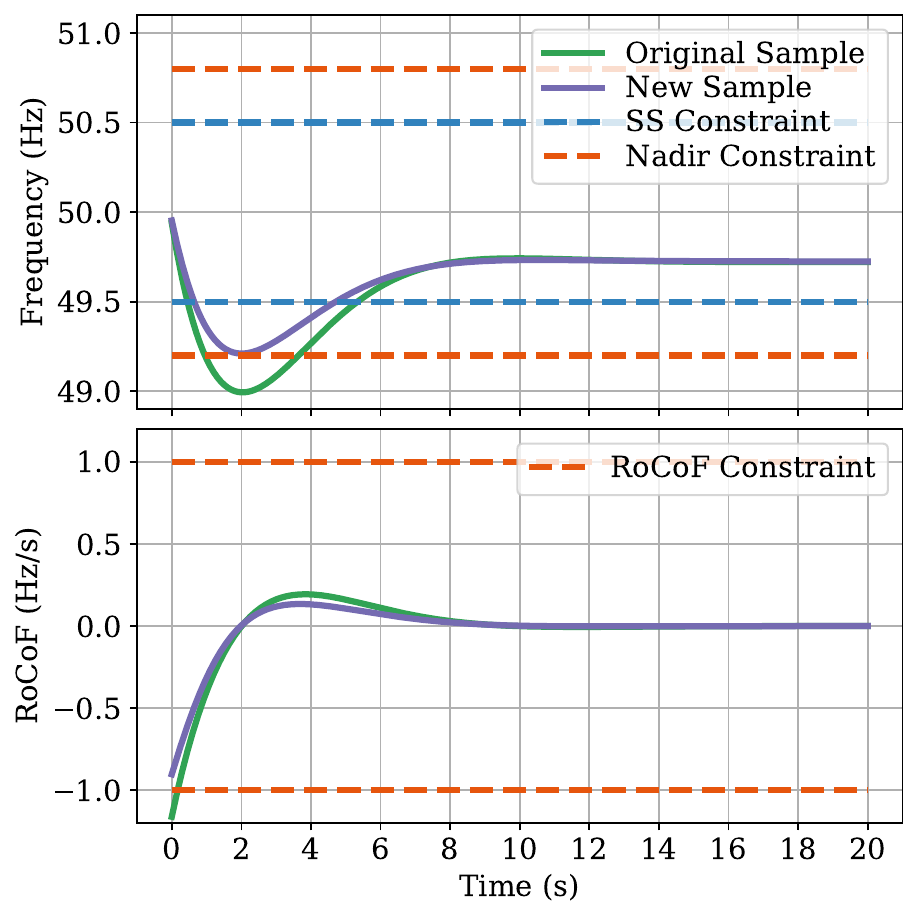}
         \caption{}
     \end{subfigure}
        \caption{Two examples of the proposed sampling method: (a). Search from `stable' sample to `unstable' sample; and (b). Search from `unstable' sample to `stable' sample.}
        \label{fig:search}
\end{figure}

Fig. \ref{fig:search} shows two examples that shift from the original sample (in green) in $\Theta_0$ to the newly generated one (in purple) in ${\Theta}$. In the first example, the frequency of the original sample is stable in terms of all three criteria. The proposed algorithm can successfully find an unstable counterpart and just terminate at the boundary of instability. In contrast, the original sample has an unstable nadir and RoCoF in the second example. The proposed algorithm can shift it to a stable counterpart.

\subsubsection{Computational Performance}

To evaluate the benefits of the proposed FMAD method, computational time and memory usage are recorded for different gradient-based methods. Two baselike methods, namely, the \textit{unrolling differentiation} and the \textit{finite difference}, are considered. A brief review of the baseline methods is given in Appendix \ref{sec:baseline}. To have a fair comparison, all the methods are evaluated on the \textbf{same and randomly-sampled} parameter $\Theta_0$ in a \textbf{batched} manner. In detail, the performance of each method is evaluated to find the gradients of samples with batch size equal to 20 in parallel and averaged over 5 runs. Memory usage is evaluated by Python package \texttt{psutil}. Table \ref{tab:comparison} also calculates the maximum absolute percentage errors of the FMAD and finite difference method with respect to the unrolling method. 

First, although the finite difference also requires one to evaluate $f(\cdot)$ four more times, its memory usage and computational time are the smallest. This is because the finite difference does not need to store any intermediate variables or calculate any gradient. However, its performance is strongly dependent on the choice of perturbation size $\epsilon$ for different parameters due to the introduction of truncation and round-off errors \cite{baydin2018automatic}. As a result, the finite difference should be avoided in real-world applications. Second, compared to unrolling AD, forward-mode AD can save 25 times of memory as it does not need to store any intermediate variables and the associated gradients on the computational graph. Meanwhile, the FMAD obtains the gradients as accurate as the unrolling AD up to machine precision. However, forward mode AD is slower than unrolling AD due to additional evaluations of $f(\cdot)$.

\begin{table*}[h]
    \footnotesize
    \centering
    \caption{Performance comparison on different gradient-based sampling methods. The errors are the maximum absolute percentage errors with respect to the solution by Unrolling AD. The performances are recorded based on 5 runs.}
    \begin{tabular}{c|c|c|c|c|c|c|c|c|c}
    \hline
         \multirow{2}{*}{\textbf{Method}} & \multicolumn{2}{c|}{\textbf{Computation Performance}} & \multicolumn{5}{c|}{\textbf{Max. Absolute Percentage Err.}} \\\cline{2-8}
         &  Memory (MB) & Time (s) & $|\Delta x(t_{ss})|$ & $|\Delta x(t_{nadir})|$ & $|\Delta x(t_{rocof})|$ & $|\Delta g_{nadir}|$ & $|\Delta g_{rocof}|$ \\\hline\hline
      Finite Difference ($\epsilon = 10^{-12}$)  & 7.41 & 0.22 & \multicolumn{3}{c|}{$\leq 10^{-14}$} & 5.29 & 3.90 \\\hline
      Finite Difference ($\epsilon = 10^{-14}$) & 7.39 & 0.22 & \multicolumn{3}{c|}{$\leq 10^{-14}$} & 10086.3977 & 5952.6488 \\\hline
      Unrolling AD & 1021.46 & 1.54 & \multicolumn{5}{c|}{N/A}\\\hline
      Forward-Mode AD & 38.76 & 3.52 & \multicolumn{5}{c|}{$\leq 10^{-14}$} \\\hline
    \end{tabular}
    \label{tab:comparison}
\end{table*}

\section{Conclusion}

This paper proposes an efficient gradient-based sampling method to construct a dataset that can be used to train frequency stability constraints in power system operation. Gradients with respect to the stationary parameters are found by solving an augmented ODE via the forward-mode automatic differentiation. Compared to the unrolling method, the simulation results demonstrate that the proposed algorithm is more memory-friendly to find the critical boundary between stable and unstable samples, with a slight trade-off on computational time.

\appendix

\subsection{Derivation of $\dot{\omega}(t=0^+)$ in \eqref{eq:closed_loop}}\label{sec:initial_condition}

For the case of $K = 0$, let $M=M_0$ and $q=\omega = 0$ in \eqref{eq:swing_equation}, it can be obtained that $\dot{\omega}(t=0^+) = \frac{\Delta P}{M_0}$. For the case of $K\neq 0$, plugging \eqref{eq:state_feedback} into \eqref{eq:swing_equation}, it gives
\begin{equation}
    K_{12}\dot{\omega}^2 - M_0\dot{\omega} + \Delta P = 0
\end{equation}
in which $\dot{\omega}$ can be solved as
\begin{equation}
    \dot{w} = \frac{g(K_{12})}{h(K_{12})}
\end{equation}
with $g(K_{12}) = M_0 \pm \sqrt{M_0^2 - 4K_{12}\Delta P}$ and $h(K_{12}) = 2K_{12}$.

Since $\lim_{K_2\rightarrow0}g(K_2) = \lim_{K_2\rightarrow 0}h(K_2) = 0$, the L'Hopital's rule gives that
\begin{equation}
    \lim_{K_2\rightarrow0}\frac{g(K_{12})}{h(K_{12})} = \lim_{K_2\rightarrow0}\frac{g'(K_{12})}{h'(K_{12})} = \pm \left(-\frac{\Delta P}{M_0}\right)
\end{equation}

Therefore, only the minus sign is feasible when $K_2\rightarrow 0$. 

\subsection{Derivation of the Augmented ODE \eqref{eq:augment_ode}}\label{sec:derive_fmad}

Starting from the \eqref{eq:ode}, the time domain solution can be written as
\begin{equation}
    x(t_1) = x_0 + \int_{t=0}^{t_1}f(x(t),\theta) dt
\end{equation}

Do partial derivative on $\theta$ on each side,
\begin{equation}
    \frac{\partial x(t_1)}{\partial \theta}=\frac{\partial x_0}{\partial \theta}+\int_{t=0}^{t_1} \frac{\partial f}{\partial x} \frac{\partial x}{\partial \theta}+\frac{\partial f}{\partial \theta} d t
\end{equation}

Multiply the unit vector $e_i$ on each side,
\begin{equation}
    \underbrace{\frac{\partial x(t_1)}{\partial \theta}e_i}_{\hat{x}_i(t_1)} = \underbrace{\frac{\partial x_0}{\partial \theta}e_i}_{\hat{x}_i(t=0^+)} + \int_{t=0}^{t_1} \frac{\partial f}{\partial x} \underbrace{\frac{\partial x}{\partial \theta}e_i}_{\hat{x}_i(t)}+\frac{\partial f}{\partial \theta}e_i d t
\end{equation}
which can be considered as the solution of the following ODE
\begin{equation}
    \frac{d \hat{x}_i(t)}{d t} = \frac{\partial f}{\partial x} \hat{x}_i(t) + \frac{\partial f}{\partial \theta}e_i, \quad \hat{x}_i(t=0^+) = \frac{\partial x_0}{\partial \theta}e_i
\end{equation}

For $i=1,3,4$, $\frac{\partial x_0}{\partial \theta}e_i = 0$. For $i=2$, referring to \eqref{eq:init_omega_dot}, $\frac{\partial x_0}{\partial \theta}e_2 = \left[0, \frac{\partial \dot{\omega}(t=0^+)}{\partial K_{12}}\right]^T$. 

\subsection{Baseline Methods}\label{sec:baseline}

\subsubsection{Unrolling Differentiation}

The Euler's method for ODE can be viewed as a function whose input is the ODE's initial values and parameters. This function is composed of elementary operations whose derivatives are known. Therefore, if all steps of the computations over $T$ are recorded on the computational graph, a direct call to the automatic differentiation package such as \texttt{PyTorch} can be used to find the gradients. To perform the parallel computation, \texttt{torch.func.vmap} is implemented. 

\subsubsection{Finite Difference}

Let $e_i$ be the $i$-th unit vector whose entry $i$ equals 1 and the remaining equals 0. Let $x(t', \theta)$ be the solution to \eqref{eq:ode} at time $t$ with parameter $\theta$. Define $\theta_{pi} = \theta + \epsilon \cdot e_i$ as the perturbed version of $\theta$ and the associated solution at time $t'$ is $x(t',\theta_{pi})$. $\epsilon$ is a small number. The gradient with respect to $\theta_i$ is approximated as
\begin{equation}
    \frac{\partial x(t',\theta)}{\partial \theta_i} \approx \frac{x(t', \theta_{pi}) - x(t', \theta)}{\epsilon}
\end{equation}

\bibliographystyle{IEEEtran}
\bibliography{IEEEabrv,Reference.bib}

\end{document}